\documentclass[letterpaper]{article}
\usepackage[preprint]{aaai2027}
\usepackage[hyphens]{url}
\usepackage{graphicx}
\usepackage{natbib}
\usepackage{caption}
\usepackage{amsmath,amssymb}
\usepackage{booktabs}
\title{Can LLM Agents Coordinate Without Negotiation?\\Evidence from Decentralized Goal Assignment}
\author{
    Deborah Sinishaw\thanks{D. Sinishaw is with the Department of Electrical and Computer Engineering, McGill University (e-mail: \texttt{deborah.sinishaw@mail.mcgill.ca}). E. Meriaux is with the School of Computer Science, McGill University (e-mail: \texttt{edwin.meriaux@mail.mcgill.ca}). G. Dudek is with the School of Computer Science, McGill University (e-mail: \texttt{gregory.dudek@mcgill.ca}). All authors are with the Centre for Intelligent Machines (CIM), McGill University.},
    Edwin Meriaux,
    Gregory Dudek
}
\affiliations{}

\begin{document}
\maketitle

\begin{abstract}
Independent agents sometimes need to make compatible choices without negotiating. We study this problem in decentralized LLM-based goal assignment, where agents have a shared description of a grid world but choose goals individually. After an initial exchange of goal rankings, agents facing the same conflict independently nominate a winner using identical language models and a common-interest matching game. We measure agreement on the first attempt, the quality of the selected outcome relative to a Nash benchmark, and the consequences for task completion. When first-round choices disagree, the simulation permits one retry and then a random tie-break, rather than open-ended negotiation. We evaluate seven locally served models on 100 seeded $20\times20$ maps, with five trials per map. Five of the seven models agree on more than 90\% of conflicts in the first round, and the two Gemma~4 models agree on 96.6\% and 99.0\%. Relative to paired lowest-index controls using the same initial rankings, this procedure reduces mean makespan for every model, and the two Gemma~4 models finish within 0.6 steps of the optimal assignment on average. The results distinguish tacit agreement from overall assignment quality and suggest that repeated negotiation may be unnecessary for most conflicts in this setting.
\end{abstract}

\section{Introduction}
How much interaction is necessary for a group of agents to make compatible decisions? In a decentralized task, agents can have the same information yet independently select actions that interfere with one another. They may negotiate until they agree, apply a fixed priority rule, or independently choose a resolution they expect their teammates to choose as well. The last option, \emph{tacit coordination}, is attractive when repeated communication is undesirable, but it raises two separate questions: do the agents agree, and is the agreement useful to the team?

We investigate these questions in multi-robot goal assignment. Each robot must reach a different goal in a fully observable grid world. Classical assignment methods can select a one-to-one matching centrally~\citep{kuhn1955hungarian}, and joint target assignment and path finding have also been studied~\citep{ma2016tapf}.

\begin{figure}[h!]
    \centering
    \includegraphics[
        width=0.84\columnwidth,
        height=0.84\textheight,
        keepaspectratio
    ]{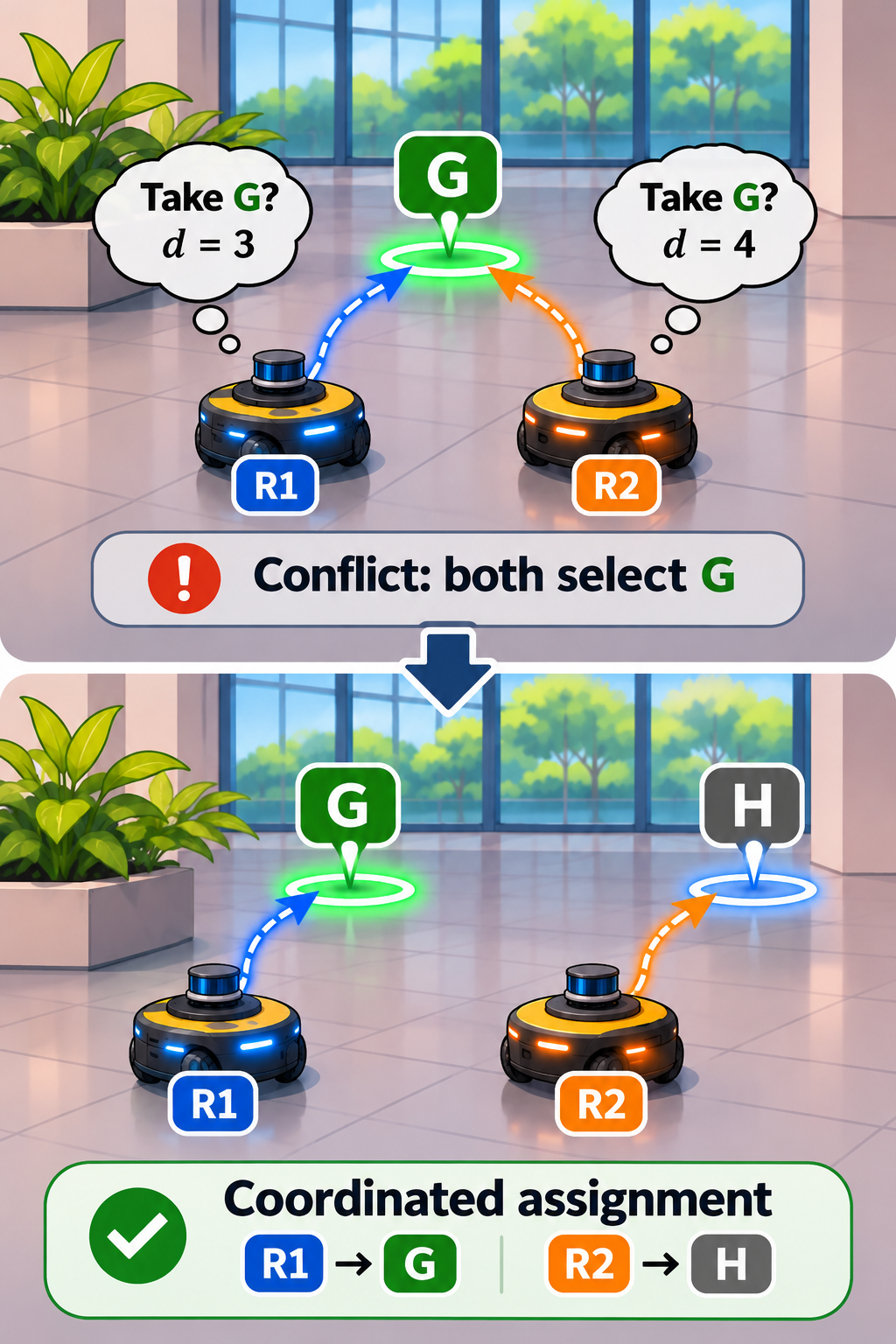}
    \caption{Illustration of decentralized goal assignment.
    Initially, both robots independently select goal $G$,
    creating a conflict. Through independent conflict-resolution
    decisions, robot $R_1$ retains goal $G$, while robot $R_2$
    switches to goal $H$, yielding a conflict-free assignment
    without negotiation.}
    \label{fig:tacit_coordination}
\end{figure}
Our concern in this work is narrower: what happens when independent agents make competing selections and must settle them without a negotiation process or with highly limited communications?

We build on the rank-once approach of \citet{ismayilov2025decentralized}: independent LLMs rank the goals, the rankings are exchanged once, and conflicts are settled by a lowest-index priority rule. That rule provides a reproducible baseline, but its selection is independent of the costs imposed on agents that yield. LLMs have also been evaluated in repeated games~\citep{akata2023playing} and game-theoretic reasoning benchmarks~\citep{fan2024rational,duan2024gtbench}. Of particular relevance, \citet{sinishaw2026nash} find strong coordination of agents in abstract matching games. Schelling's focal points offer a possible explanation for reaching a common choice without negotiation~\citep{schelling1960strategy}. Here, each choice also changes the downstream navigation task.

In decentralized multi-agent goal assignment, robots independently select destinations, but their preferred goals may overlap. Resolving these conflicts requires deciding which robot should retain a contested goal and which should pursue an alternative. We investigate whether identical LLM agents, given shared environmental information and agent--goal distances, can independently reach consistent and beneficial assignment decisions without negotiation. We formulate these conflicts as common-interest matching games and evaluate how frequently agents agree on the first attempt, the quality of their decisions, and the resulting impact on team makespan. We further compare this approach against the fixed-priority conflict resolution of \citet{ismayilov2025decentralized}, using identical initial goal rankings to isolate the effect of conflict resolution.

\begin{figure*}[t]
\centering
\includegraphics[width=\textwidth]{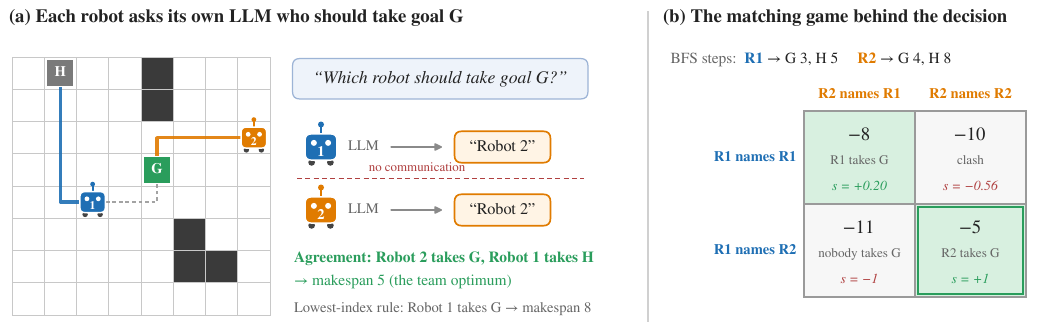}
\caption{Tacit conflict resolution and its local game. (a)~R1 and R2 both prefer goal $G$. Independently prompted copies of the same LLM both nominate R2, giving R2 goal $G$ and R1 goal $H$ for a makespan of 5; the lowest-index rule assigns $G$ to R1 and yields 8 steps in this example. (b)~The common payoff matrix is the negative local makespan, using BFS distances and conflict penalties $K=6$, $D=3$. Agreeing on R2 is the best outcome ($s=1$); $N=-8.75$ is the Nash reference used to normalize scores.}
\label{fig:key}
\end{figure*}

\section{Problem Formulation and Method}

We consider decentralized multi-agent goal assignment and
formulate conflicts between agents as common-interest games where prolonged communication is to be avoided.
We first define the assignment problem and the information
available to each agent, then introduce the game used to
evaluate competing decisions, and finally describe the
independent decision-making and fallback procedure.

\paragraph{Decentralized goal assignment.}
We consider $n$ agents and $n$ goals in a fully observable
$L\times L$ grid world containing obstacles. Each agent must
reach a distinct goal, and the objective is to minimize the
team makespan, defined as the number of timesteps until all
agents have reached their assigned goals. Agents move one
cell per timestep along breadth-first-search (BFS) shortest
paths, without using the LLM for individual movement actions.

Following \citet{ismayilov2025decentralized}, each agent
initially receives the environment description, agent and
goal positions, obstacle locations, and the agent--goal BFS
distance table. Image-capable models additionally receive
the grid image. Each agent independently produces a complete
ranking of the goals, and these rankings are exchanged once.
Conflicts arise when multiple agents select the same goal
or attempt to enter the same cell. The original method
resolves these conflicts by prioritizing the lowest-index
agent. We instead formulate each conflict as a game in
which the participating agents independently select a
common resolution.

\paragraph{Common-interest matching game.}
Consider a conflict involving a set of agents $P$ competing
for the same goal or cell $G$. We define a game in which
each participating agent $i\in P$ selects an action
$c_i\in P$, identifying the agent it believes should
receive $G$. If all agents select the same agent $j$,
they coordinate: $j$ takes $G$, while the others pursue
their next-ranked goals or wait, depending on the conflict.

To evaluate each outcome, let $a_i$ denote the cost to
agent $i$ of taking $G$, and $b_i$ the cost of yielding.
For goal conflicts, these correspond to BFS distances
to the contested goal and the next-ranked available goal.
For movement conflicts, they represent the remaining
travel cost after moving or waiting, respectively.

The game is common-interest: all participating agents
receive the same payoff, defined as the negative local
makespan. We additionally introduce penalties $K=0.75L$
for conflicting claims and $D=0.375L$ when no agent
takes the contested location. These penalties discourage
disagreement and are part of the game payoff rather than
physical movement delays. The shared payoff is

\begin{equation}
u=\begin{cases}
-\max\!\big(a_j,\max_{i\ne j}b_i\big),
    & \text{all name }j,\\[2pt]
-\max\!\big(K+\max_{i\in T}a_i,
           \max_{i\notin T}b_i\big),
    & T\ne\emptyset,\\[2pt]
-\big(\max_i b_i+D\big),
    & T=\emptyset,
\end{cases}
\label{eq:payoff}
\end{equation}

where, for disagreement,
$T=\{i\in P:c_i=i\}$ is the set of agents that nominate
themselves; empty maxima are omitted. The latter two
cases apply only when the agents do not unanimously
select the same winner.

Thus, the game distinguishes between agreement on a
common decision and the quality of that decision.
Different agreements can have different payoffs, with
the highest-payoff agreement minimizing the predicted
local makespan. Figure~\ref{fig:key} illustrates a
case in which assigning the contested goal to the
higher-index agent produces a better outcome than
the original fixed-priority rule.

\paragraph{Independent decisions and bounded fallback.}
Each agent involved in a conflict independently queries
its own copy of the same LLM. All agents receive the
same conflict description, candidate costs, and payoff
information, differing only in the identification of
the responding agent. The prompts explicitly state
that the other agents are identical models making
decisions independently, without communication.
Each model returns a JSON response identifying
its preferred agent, together with a brief justification.
For conflicts involving more than two agents, the
prompt provides the payoffs of unanimous decisions
and the disagreement rule rather than enumerating
every possible combination of responses.

If all agents independently select the same winner,
the conflict is resolved immediately. Otherwise, the
agents wait one timestep and repeat the decision once
with the candidate order reversed. If disagreement
persists, a seeded uniform random selection among the
competing agents resolves the conflict after a second
waiting step. Consequently, each conflict requires
at most two independent LLM decision rounds before
the fallback is applied, avoiding open-ended negotiation.

We evaluate first-round decisions separately from the
fallback procedure to measure the agents' ability to
coordinate independently. For comparison, we retain
the original lowest-index rule as a paired control,
using identical initial goal rankings, movement
procedures, and re-planning behavior. This isolates
the effect of the conflict-resolution mechanism
on final task performance.

\begin{table*}[t]
\centering
\small
\setlength{\tabcolsep}{5pt}
\begin{tabular}{@{}llccccccccc@{}}
\toprule
 & & \multicolumn{2}{c}{Mean makespan $\downarrow$} & \multicolumn{2}{c}{Optimal trials $\uparrow$} & \multicolumn{2}{c}{Failed trials $\downarrow$} & \multicolumn{3}{c}{Per-conflict LLM decisions} \\
\cmidrule(lr){3-4}\cmidrule(lr){5-6}\cmidrule(lr){7-8}\cmidrule(l){9-11}
Model & Ranking input & rule & LLM & rule & LLM & rule & LLM & Round-1 agree & Mean $s$ & Random draws \\
\midrule
Gemma 4 E4B    & image + text & 14.37 & \textbf{13.74} & 362 & 383 & 4 & 0 & 96.6\% & 0.94 & 0.3\% \\
Gemma 4 12B    & image + text & 14.06 & \textbf{13.78} & 375 & 394 & 3 & 0 & 99.0\% & 0.98 & 0.0\% \\
Phi-4 14B      & text         & 15.26 & \textbf{14.26} & 284 & 324 & 8 & 4 & 90.2\% & 0.80 & 0.6\% \\
Ministral 3 8B & image + text & 16.21 & \textbf{14.77} & 255 & 297 & 7 & 1 & 90.9\% & 0.80 & 0.9\% \\
Qwen2.5-VL 7B  & image + text & 18.13 & \textbf{15.37} & 144 & 247 & 9 & 0 & 95.9\% & 0.94 & 0.9\% \\
Llama 3.2 3B   & text         & 18.16 & \textbf{17.07} & 145 & 135 & 5 & 4 & 70.7\% & 0.48 & 9.3\% \\
Llama 3.1 8B   & text         & 18.26 & \textbf{17.36} & 126 & 152 & 1 & 0 & 60.6\% & 0.29 & 16.3\% \\
\midrule
\multicolumn{11}{@{}l}{Optimal assignment reference: 13.17. Greedy (no LLM): 16.11, 205 optimal and 10 failed trials.} \\
\bottomrule
\end{tabular}
\caption{Results for 100 maps $\times$ 5 trials per model. Paired rule controls use the same initial rankings. Makespan means exclude failed trials (100-step cap). Decision-level columns aggregate conflicts within each model's LLM trials; the score is defined in Eq.~\ref{eq:score}.}
\label{tab:main}
\end{table*}

\section{Measuring Coordination}

A successful first-round match says whether agents agree, but not whether they agree on a good action. We therefore record the \emph{first-round agreement rate}, the fraction of conflicts that require the \emph{random fallback}, and a score for each first-round outcome. Following \citet{sinishaw2026nash}, let $u$ be its team payoff, $O$ and $W$ the best and worst possible payoffs, and $N$ a Nash benchmark. We use the worst Nash equilibrium, allowing asymmetric mixed strategies because agents have different travel costs:
\begin{equation}
 s=\begin{cases}
 \dfrac{u-N}{O-N}, & u\ge N,\\[7pt]
 \dfrac{u-N}{N-W}, & u<N,
 \end{cases}\quad s\in[-1,1].
 \label{eq:score}
\end{equation}
Here $s=1$ denotes the best outcome, $s=-1$ the worst, and $s=0$ the chosen Nash reference. If $N=O$, the score is undefined and that conflict is omitted from the score mean. Failed first-round coordination is still included when computing decision scores. This metric evaluates the quality of decisions relative to a game-theoretic reference; it does not measure communication cost directly.

\section{Experimental Design}

We generate 100 seeded $20\times20$ maps with 2--6 agents, an equal number of goals, and 15--30 obstacles placed away from agent and goal cells, using the generator settings of \citet{ismayilov2025decentralized}. The earlier maps were not seeded and are unavailable, so ours are not the same instances. We run five fresh trials per map and model, totaling 500 trials per model and 3,500 LLM trials overall.

Seven open-weight models are run locally with Ollama: Gemma~4 E4B, Gemma~4 12B (thinking disabled), Qwen2.5-VL 7B, and Ministral~3 8B use image and text for goal rankings; Phi-4 14B, Llama~3.1 8B, and Llama~3.2 3B use text only. Every conflict prompt is text-only. Calls use temperature 0 and an 8,192-token context, and the order of agents in conflict prompts alternates across trials. Every LLM run has a paired \emph{rule} control that reuses its exact initial goal rankings and movement/re-planning code but chooses the lowest-index agent at each conflict. This lowest-index rule is the conflict-resolution step of \citet{ismayilov2025decentralized} and serves as our baseline.

For task-level evaluation, we report successful-trial mean makespan, optimal outcomes, and failures at the 100-step cap. The optimal assignment reference is computed by brute-force minimization of the maximum agent--goal BFS distance over one-to-one assignments; a greedy baseline uses nearest-goal BFS distances with lowest-index tie-breaking. Means exclude failed trials. We additionally compare with the earlier study using each dataset's own greedy-to-optimal gap, rather than comparing absolute makespans across different maps.

\section{Results}

\paragraph{How often do agents agree?}
Across the seven models, the LLM trials generated 8,511 conflicts; 98\% were goal conflicts, and 82\% involved two agents. Five of seven models agreed on more than 90\% of conflicts in the first round (Table~\ref{tab:main}). Gemma~4 E4B and 12B reached 96.6\% and 99.0\% agreement, with mean scores $s=0.94$ and $0.98$; their random-fallback rates were 0.3\% and 0.0\%. Phi-4, Ministral~3, and Qwen2.5-VL also exceeded 90\% agreement, whereas the two Llama models reached 70.7\% and 60.6\% and needed random resolution in 9.3\% and 16.3\% of conflicts. Thus tacit coordination depends strongly on the model; it is not a universal property of LLM agents.

\paragraph{Does it help the team?}
For every model, the LLM conflict decider has a lower mean makespan than its paired lowest-index rule, improving by 0.28--2.76 steps. Across models the number of failed trials decreases from 37 to 9. These are average gains, not per-trial dominance: for each model, some paired trials finish later with the LLM decider, and Llama~3.2 3B produces fewer optimal trials (135 rather than 145) despite its lower mean makespan. The Gemma~4 models finish in 13.74 and 13.78 steps on average, compared with the 13.17 optimal-assignment reference, and close approximately 81\% and 79\% of the gap between the greedy (16.11) and optimal references. On its \emph{different} maps, the rank-once GPT-4.1 configuration of \citet{ismayilov2025decentralized} achieved 15.50 steps, compared with 17.93 greedy and 13.93 optimal, closing approximately 61\% of that dataset's gap. These cross-study figures give context, not a paired head-to-head comparison.

\paragraph{What do decision metrics explain?}
High agreement does not imply good initial rankings. Qwen2.5-VL agrees on 95.9\% of conflicts but generates 1,497 conflicts, more than twice the 653--668 reported for the Gemma~4 models; its makespan is correspondingly higher. Both Llama models remain worse than the greedy baseline even with LLM resolution. Agreement, payoff quality, and conflict frequency therefore describe distinct contributors to the final makespan. The per-conflict measurements offer more diagnostic information than task completion alone, although seven model-level observations do not establish a general predictive relationship.

\section{Discussion and Conclusion}

Our results demonstrate that independently queried copies
of the same LLM can consistently make beneficial
goal-assignment decisions without negotiation.
Across 8,511 conflicts, five of the seven evaluated models
achieved first-round agreement rates above 90\%.
In particular, Gemma~4 E4B and Gemma~4 12B achieved
96.6\% and 99.0\% agreement, respectively, with
corresponding mean game-theoretic coordination scores
of 0.94 and 0.98. These results extend the findings
of \citet{sinishaw2026nash} from abstract game-theoretic
settings to decentralized goal assignment, where
the agents' decisions directly influence the
performance of the team.

Importantly, this coordination translated into
improvements in task performance. All seven models
achieved lower mean makespans with the proposed
conflict-resolution procedure than with the fixed-priority
rule of \citet{ismayilov2025decentralized}, using
identical initial goal rankings. The two
Gemma~4 models achieved mean makespans of 13.74 and
13.78 steps, compared with the optimal assignment
reference of 13.17, and completed all 500 trials
successfully. These results show that the ability to
reach consistent conflict-resolution decisions can
produce substantial improvements in collective
task performance.

A particularly promising implication is the potential
for decentralized agents to operate with limited
communication. Given shared environmental information
and an initial exchange of goal rankings, agents can
independently select compatible resolutions without
negotiating each conflict. When disagreements occur,
the proposed procedure permits one additional independent
decision before applying a random tie-break, avoiding
open-ended negotiation. For the two Gemma~4 models,
this final fallback was needed in at most 0.3\% of
conflicts. The high first-round agreement rates suggest
that, in environments where communication is intermittent
or restricted, agents could frequently make compatible
goal-assignment decisions using common information
rather than repeated message exchanges. When agreement is this reliable, agents could exchange rankings once and then resolve conflicts with no further messages, accepting occasional failures; this may suit settings with severely limited communication.

Finally, the experiments highlight the distinction
between initial goal-selection quality and conflict
coordination. Models that coordinate well do not
necessarily produce the best initial goal rankings,
as illustrated by Qwen2.5-VL, which achieved 95.9\%
first-round agreement but generated substantially
more conflicts than the Gemma~4 models. Evaluating
agreement, coordination quality, and final makespan
together therefore provides a more complete picture
of collective decision-making. Overall, our findings
show that effective decentralized goal assignment
can be achieved through independent LLM decisions
and bounded conflict resolution, with most conflicts
requiring no iterative negotiation.

\bibliography{references}

\section*{Use of AI Tools}
Figure~\ref{fig:tacit_coordination} was created with a generative AI image tool. An AI assistant was also used to help edit the text and to write the plotting code for Figure~\ref{fig:key}. The authors reviewed and verified all content and take full responsibility for it.
\end{document}